# Molecular potentials for 2D molybdenum disulphide: transferability and performance

Marcin Maździarz*

*Institute of Fundamental Technological Research Polish Academy of Sciences, Pawińskiego 5B, 02-106 Warsaw, Poland*

**Abstract**

An ability of different molecular potentials to reproduce the properties of 2D molybdenum disulphide polymorphs is examined. Structural and mechanical properties, as well as phonon dispersion of the 2H, 1T and 1T' single-layer MoS$_2$ (SL MoS$_2$) phases, were obtained using density functional theory (DFT) and molecular statics calculations (MS) with Stillinger-Weber, REBO, SNAP, and ReaxFF potentials. Quantitative systematic comparison and discussion of the results obtained are reported.

## 1. Introduction

Group 6 transition metal dichalcogenide (G6-TMDs) two-dimensional (2D) nanomaterials [1], and especially single-layer molybdenum disulphide (SL MoS$_2$), are probably the second most studied 2D materials following graphene [2]. The major disadvantages of graphene are the lack of a band gap in the electronic spectrum, its susceptibility to oxidative environments, and that it has some toxic properties. That is why scientists and engineers, beyond ordinary human curiosity, have begun to look for materials free of these deficiencies [3, 4].

Both synthetic and natural bulk transition metal dichalcogenides have layered structures with two primary distinguished allotropic forms, 2H and 3R, belonging to the hexagonal crystal family, but differing in a sequence of arrangement. Strong triple layers of metal-sulphur-metal are weakly bounded by the van der Waals forces, similar to graphene in graphite [1].

Three polymorphs of single-layer molybdenum disulphide have been synthesised, namely the most thermodynamically stable semiconducting 1H-MoS$_2$, semimetallic 1T'-MoS$_2$, and metastable metallic 1T-MoS$_2$ [5]. In 1H-MoS$_2$

*Corresponding author

*Email address:* mmazdz@ippt.pan.pl (Marcin Maździarz)

structural phase, the S and Mo atoms are stacked in an A-B-A order, the 1T-$MoS_2$ dynamically unstable phase has an A-B-C stacking, whereas 1T'-$MoS_2$ phase is a disturbed 1T-$MoS_2$ phase [6].

The most accurate methods of solid state physics are based on quantum mechanics, unfortunately, with the accuracy of the methods their cost increases. The number of atoms and the number of timesteps that can be analysed with the first-principles method using either energy minimisation or *ab initio* molecular dynamics (AIMD) is highly limited. For typical computational resources currently available, the use of these methods is limited to several hundreds of atoms for less than about 10 picoseconds. These restrictions justify the need for more approximate methods, such as molecular methods [7].

In general, there is a lack of perfectly transferable interatomic potential that would work with the various materials and systems we are interested in. Some are more transferable, others less [8]. It depends on the physics behind them, the mathematical flexibility of the model capable of describing the multimodal potential energy surface (PES) and the quality of the fitting process and, of course, on the "difficulty" of the material [7].

According to the author's best knowledge, there are no publications where the performance of different molecular potentials for molybdenum disulphide is analysed for all phases of SL $MoS_2$, there are only partial comparisons, and so in [9] the results for 1H and 1T phases for potentials Stillinger-Weber, REBO and ReaxFF are only compared between each other. In [10] the geometric parameters and mechanical properties of 1H phase obtained from Stillinger-Weber and REBO potentials are compared with density functional theory (DFT) calculations. Thermal transport properties in 1H phase from molecular dynamics using Stillinger-Weber and REBO potentials were obtained in [11].

A partial comparison of different potentials for the 1H phase SL $MoS_2$ can be found in papers where new parametrisations are presented, e.g. [12, 13, 14]. There are also publications where using molecular simulations the Authors try to determine certain SL $MoS_2$ properties that were not taken into account during the parametrization of potential, e.g. [15, 16, 17, 18].

The paper is organised as follows. Following the above Introduction 1, Section 2.1 presents the computational methodology used in *Ab initio* calculations of analysed structures and Section 2.2 describes the computational methodology used in molecular calculations and molecular potentials examined: four Stillinger-Weber (SW) potentials [19], the reactive many-body (REBO) potential, the spectral neighbour analysis potential (SNAP) and the reactive force-field (ReaxFF). Section 3 presents the structural and mechanical properties of SL $MoS_2$ and phonon spectra obtained from the *Ab initio* and molecular calculations and evaluates the quality of the analysed potentials. Last Section 4 summaries and concludes the results obtained.

## 2. Computational methodology

Analysing the available literature concerning phases of SL $MoS_2$, it is not feasible to find all structural, mechanical and phonon data obtained in one con-sistent way. The availability of experimental data is actually limited to phase 1H only and therefore we must use *ab initio* calculations. Unfortunately, also *ab initio* calculations, most often DFT, differ in the calculation methodology, i.e., they use different functional bases, different pseudopotential or exchange-correlation (XC) functionals, and such a parameter as cohesive energy is not accessible at all. For this reason structural and mechanical data: lattice parameters, average cohesive energy, average bond length, average height, 2D elastic constants as well as phonon data are determined using a single consistent first-principle approach as described in the next Section 2.1. These data will be further considered as reference data and marked as $\text{Value}^{\text{DFT}}$. Then the same data were determined, as described in Section 2.2, using the analysed molecular potentials 2.2.1 and will be marked as $\text{Value}^{\text{potential}}$. Having both data we can simply define mean absolute percentage error (MAPE):

$$\text{MAPE} = \frac{100\%}{n} \sum_{t=1}^{n} \left| \frac{\text{Value}^{\text{DFT}} - \text{Value}^{\text{potential}}}{\text{Value}^{\text{DFT}}} \right|, \quad (1)$$

that will allow us to quantify the potentials under examination. Phonons were determined only for the three best, having the lowest MAPE, molecular potentials.

### 2.1. Ab initio calculations

First-principle calculations based on density functional theory (DFT) [20, 21] within the pseudopotential plane-wave approximation (PP-PW) implemented in ABINIT [22, 23] code were performed in this work. Optimised norm-conserving Vanderbilt pseudopotentials (ONCVPSP) [24] were used to describe the interactions of ionic core and non-valence electrons. ONCVPSP pseudopotentials used were taken from PseudoDojo project [25].

To strengthen the reliability of the calculations as an exchange-correlation (XC) functional, three approximations were initially checked for their ability to reproduce the geometry of 1H-MoS$_2$: local density approximation (LDA) [26, 27], classical Perdew-Burke-Ernzerhof (PBE) generalised gradient approximation (GGA) [28] and modified Perdew-Burke-Ernzerhof GGA for solids (PBEsol) [29]. To provide access to all XC functionals used a library of exchange-correlation functionals for density functional theory, LibXC [30] was utilised.

All the computations were made by adjusting their precision, which was done by automatically setting the variables at *accuracy* level 4 (*accuracy*=4 corresponds to the default tuning of ABINIT). The *cut-off* energy consistent with ONCVPSP pseudopotentials of the plane-wave basis set was 35 Ha with $4d^5 5s^1$ valence electrons for Mo and $3s^2 3p^4$ valence electrons for S. K-PoinTs grids were generated with *kptrlen*=35.0 (grids that define a length of the smallest vector LARGER than *kptrlen*). Metallic occupation of levels with the Fermi-Dirac smearing occupation scheme and *tsmear* (Ha)=0.02 was used in all ABINIT computations.

Initial data defining unit cells of SL 1H-MoS$_2$, 1T-MoS$_2$ and 1T'-MoS$_2$ were taken from [31] and then all structures were relaxed by applying the Broyden-Fletcher-Goldfarb-Shanno minimisation scheme (BFGS) with full optimisation of cell geometry and atomic coordinates. Maximal stress tolerance (GPa) was set to $1\times10^{-4}$.

The cohesive energy of the $E_c(MoS_2)$ is calculated, taking into account stoichiometry, as the total energy $E_{total}(MoS_2)$ difference of 2D molybdenum disulphide and a single Mo atom energy $E_{iso}(Mo)$ in a sufficiently large box and a single S atom energy $E_{iso}(S)$ in a similar box [32]:

$$E_c(MoS_2) = E_{total}(MoS_2) - E_{iso}(Mo) - 2E_{iso}(S). \qquad (2)$$

The theoretical ground state elastic constants, $C_{ij}$, of all structures were identified with the metric tensor formulation of strain in density functional perturbation theory (DFPT) [33].

In order to examine the elastic (mechanical, Born) stability of all the structures, positive definiteness of the elasticity tensor was checked [34] by calculating Kelvin moduli, i.e., eigenvalues of stiffness tensor represented in *second-rank tensor* notation [35, 36].

To calculate phonons, DFPT was utilised [22, 23]. The phonon dispersion curves (for 1H-MoS$_2$ and 1T-MoS$_2$: $\Gamma$[0,0,0]-**M**[1/2,0,0]-**K**[1/3,1/3,0]-$\Gamma$[0,0,0], and for 1T'-MoS$_2$: $\Gamma$[0,0,0]-**Z**[0,1/2,0]-**C**[1/2,1/2,0]-**Y**[1/2,0,0]-$\Gamma$[0,0,0]) [37] of the analysed structures were then used to identify their dynamical stability [34, 38], complementary to elastic stability.

### 2.2. Molecular calculations

The molecular statics (MS) method (i.e. at 0 K temperature) [7, 39, 40] simulations were made using the Large-scale Atomic/Molecular Massively Parallel Simulator (LAMMPS) [41] and analysed in the Open Visualization Tool (OVITO) [42].

To get the components of the elasticity tensor, $C_{ij}$, for all pre-relaxed structures, the stress-strain method with the maximum strain amplitude of $10^{-6}$ was utilised [41, 43].

For the phonon calculations, phonoLAMMPS (LAMMPS interface for phonon calculations using Phonopy code [44]) [45] was utilised. Supercell and finite displacement approaches were used with 3x3x1 supercell of the unit cell and the atomic displacement distance of 0.01 Å. The cohesive energy, $E_c(MoS_2)$, (Eq.2) in molecular calculations is simply potential energy.

### 2.2.1. Molecular potentials

- **SW2013**[13]: the Stillinger-Weber (SW) potential fitted to an experimentally obtained phonon spectrum along the $\Gamma$-M direction for bulk 2H-MoS$_2$.

- **SW2015**[14]: the Stillinger–Weber (SW) potential derived from the valence force-field model.

- **SW2016**[46]: the Stillinger-Weber (SW) potential fitted to lattice parameters, distance between two chalcogen atoms and elastic constants for SL 1H-MoS$_2$ obtained from DFT calculations.

- **SW2017**[12, 47]: the force-matching Stillinger-Weber (SW) potential fitted to first principles forces for a training set of atomic configurations of SL 1H-MoS$_2$.

- **REBO**[48]: the reactive many-body potential (REBO) fitted to structure and energetics of Mo molecules, two-dimensional Mo structures, three-dimensional Mo crystals, small S molecules, and binary Mo-S crystal structures.

- **SNAP**[49]: the machine-learning-based spectral neighbour analysis potential (SNAP) fitted to total energies and interatomic forces in SL 1H-MoS$_2$ obtained from first-principles density-functional theory (DFT) calculations.

- **ReaxFF**[50]: the reactive force-field (ReaxFF) parameters fitted to a training set of energies, geometries, and charges derived from DFT calculations for both clusters and periodic Mo$_x$S$_y$ systems.

## 3. Results

The first step of the *ab initio* calculation was to select the exchange-correlation (XC) functional that most accurately reproduces the experimental geometry of 1H-MoS$_2$. The measured lattice constant for SL 1H-MoS$_2$ a=3.157 Å and average height (vertical separation between S atoms) $h$=3.116 Å [50], while that calculated with the local density approximation (LDA) a=3.144 Å, $h$=3.111 Å, with the classical Perdew-Burke-Ernzerhof (PBE) generalised gradient approximation (GGA) a=3.220 Å, $h$=3.121 Å and with the modified Perdew-Burke-Ernzerhof GGA for solids (PBEsol) a=3.165 Å, $h$=3.120 Å, a similar trend can also be observed in other papers [12, 51]. Once again, it was confirmed that the PBEsol is the overall best performing XC functional for identifying the structural and mechanical properties [52, 53, 4] and thus all subsequent calculations will use PBEsol XC functional.

*3.1. Structural and mechanical properties*

The basic cell for the SL 1H-MoS$_2$ polymorph is depicted in Fig.1 (*hP3* in Pearson notation, P$\bar{6}$m2–space group in Hermann–Mauguin notation, no.187–space group in the International Union of Crystallography (IUCr) notation), SL 1T-MoS$_2$ polymorph is shown in Fig.2 (*hP3*, P$\bar{3}$m1, no.164) and SL 1T'-MoS$_2$ is depicted in Fig.3 (*oP6*, P2$_1$/m, no.11), respectively [54]. Although 2D structures were studied, as is commonly practised, the 3D notation is used here. The crystallographic data for all calculated phases are additionally stored in crystallographic information files (CIFs) in Supplementary Materials Appendix A .

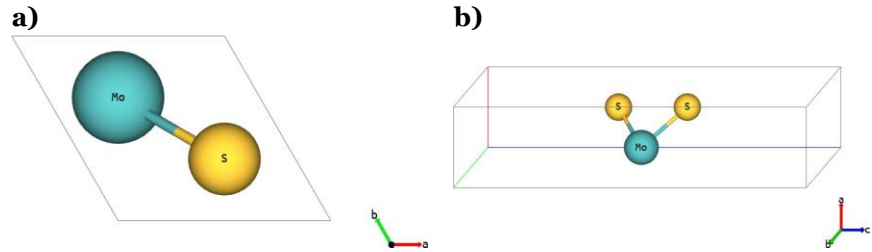

Figure 1: SL 1H-MoS$_2$—**a**) Top and **b**) 3D view.

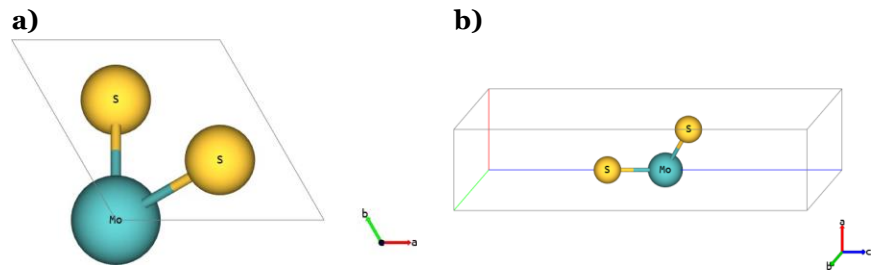

Figure 2: SL 1T-MoS$_2$—**a**) Top and **b**) 3D view.

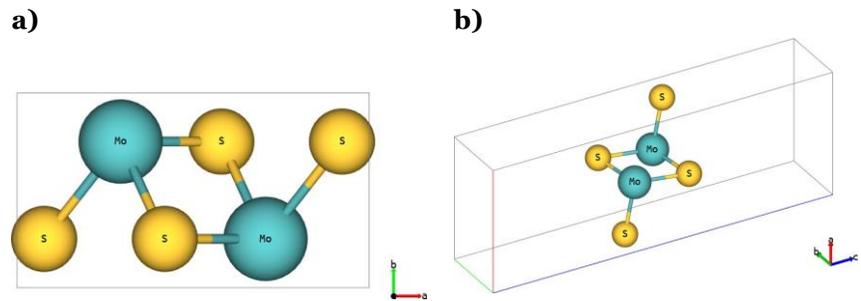

Figure 3: SL 1T'-MoS$_2$—**a**) Top and **b**) 3D view.

Determined from DFT calculations structural and mechanical properties, namely, lattice parameters, average cohesive energy, average bond length, average height, 2D elastic constants and 2D Kelvin moduli, of the three analysed SL MoS$_2$ allotropes are gathered in Tab.1. It can be seen that the calculated values match well the available experimental data [50] as well as those from other calculations [55, 56, 57]. This can be regarded as a confirmation of the correctness of the applied methodology. It is worth observing that the trend in the calculated cohesive energy matches the stability of the analysed phases and adding that all the calculated values are obtained using one consistent methodological approach. All calculated 2D Kelvin moduli for the three analysed phases are positive, which translates into mechanical stability.

Table 1: Structural and mechanical properties of SL $MoS_2$ phases from DFT calculations: lattice parameters a,b (Å), average cohesive energy $E_c$ (eV/atom), average bond length $d$ (Å), average height $h$ (Å), 2D elastic constants $C_{ij}$ (N/m) and 2D Kelvin moduli $K_i$ (N/m)

| Polymorph | 1H | | | 1T | | | 1T′ | | |
|---|---|---|---|---|---|---|---|---|---|
| Source | Present | Exp. | DFT | Present | Exp. | DFT | Present | Exp. | DFT |
| a | 3.165 | 3.157[a] | 3.183[b] | 3.194 | | 3.179[b] | 5.751 | | 5.717[b] |
| b | 3.165 | 3.157[a] | 3.183[b] | 3.194 | | 3.176[b] | 3.177 | | 3.179[b] |
| $-E_c$ | 5.64 | | 5.35[a] | 5.52 | | | 5.56 | | |
| $d_{Mo-S}$ | 2.403 | 2.38[a] | 2.43[a] | 2.422 | | 2.430[c] | 2.415[‡] | | |
| $h_{S-S}$ | 3.120 | 3.116[a] | 3.11[a] | 3.142 | | 3.184[c] | 3.364 | | |
| $C_{11}$ | 126.5 | | 127.2[d] | 84.1 | | 103.8[d] | 68.1 | | 94.0[d] |
| $C_{22}$ | 126.5 | | 127.2[d] | 84.1 | | 103.8[d] | 78.9 | | 119.2[d] |
| $C_{12}$ | 28.5 | | 25.8[d] | 5.0 | | -2.5[d] | 18.2 | | 17.2[d] |
| $C_{44}$ | 49.0 | | 51.0[d] | 39.6 | | 52.8[d] | 43.2 | | 37.5[d] |
| $K_I$ | 155.0 | | | 89.1 | | | 90.9 | | |
| $K_{II}$ | 98.0 | | | 79.1 | | | 56.1 | | |
| $K_{III}$ | 98.0 | | | 79.1 | | | 86.4 | | |

[a] Ref.[50], [b] Ref.[55], [c] Ref.[56], [d] Ref.[57]
[‡] average first-neighbour bond lengths calculated with *cutoff* radius=3.5 and number of histogram bins=50

Calculated with the use of molecular statics and different molecular potentials twelve structural and mechanical properties, namely, lattice parameters, average cohesive energy, average bond length, average height, 2D elastic constants and 2D Kelvin moduli, of the SL 1H-$MoS_2$ phase are collected in Tab.2. The results obtained are then compared with those from DFT and quantified by calculating the mean absolute percentage error (MAPE) using the Eq.1. What follows from the results obtained. Overall, analysing the $MAPE_{1H}$ for 1H-$MoS_2$, the three most accurate potentials are: SW2017, SNAP, REBO, the least: ReaxFF and SW2015. A detailed look shows that only two potentials correctly reproduce cohesive energy. Mechanical stability is correctly reproduced by all potentials, i.e., $K_i > 0$. Potential ReaxFF catastrophically badly reproduces mechanical properties of 1H-$MoS_2$, even the symmetry of the elasticity tensor is not correct.

Table 2: Structural and mechanical properties of SL 1H-MoS$_2$ from molecular calculations: lattice parameters a,b (Å), average cohesive energy $E_c$ (eV/atom), average bond length $d$ (Å), average height $h$ (Å), 2D elastic constants $C_{ij}$ (N/m), 2D Kelvin moduli $K_i$ (N/m), mean absolute percentage error (MAPE) (%)

| Method | DFT | SW2013 | SW2015 | SW2016 | SW2017 | REBO | SNAP | ReaxFF |
|---|---|---|---|---|---|---|---|---|
| a | 3.165 | 3.062 | 3.117 | 3.174 | 3.196 | 3.168 | 3.139 | 3.186 |
| b | 3.165 | 3.062 | 3.117 | 3.174 | 3.196 | 3.168 | 3.139 | 3.186 |
| $-E_c$ | 5.64 | 3.00 | 0.62 | 1.84 | 5.11 | 7.16 | 2.28 | 5.05 |
| $d_{Mo-S}$ | 2.403 | 2.399 | 2.382 | 2.515 | 2.441 | 2.445 | 2.392 | 2.431 |
| $h_{S-S}$ | 3.120 | 4.223 | 4.257 | 4.032 | 3.194 | 3.242 | 3.124 | 3.183 |
| $C_{11}$ | 126.5 | 103.9 | 45.8 | 90.0 | 118.9 | 154.4 | 140.3 | 237.3 |
| $C_{22}$ | 126.5 | 103.9 | 45.8 | 90.0 | 118.9 | 154.4 | 140.3 | 262.4 |
| $C_{12}$ | 28.5 | 33.4 | 8.0 | 30.1 | 40.9 | 45.8 | 35.7 | 121.2 |
| $C_{44}$ | 49.0 | 35.2 | 18.9 | 30.0 | 39.0 | 54.3 | 52.3 | 71.2 |
| $K_I$ | 155.0 | 137.3 | 53.8 | 120.1 | 159.8 | 200.2 | 176.0 | 370.4 |
| $K_{II}$ | 98.0 | 70.5 | 37.8 | 59.9 | 78.0 | 108.6 | 104.6 | 129.3 |
| $K_{III}$ | 98.0 | 70.4 | 37.8 | 60.0 | 78.0 | 108.6 | 104.6 | 142.4 |
| MAPE $_{1H}$ |  | 19.797 | 48.204 | 25.342 | 11.263 | 16.602 | 11.886 | 66.398 |

Computed twelve structural and mechanical properties of the SL 1T-MoS$_2$ phase are summarised in Tab.3. In general, analysing the MAPE $_{1T}$ for 1T-MoS$_2$, the three most accurate potentials are: SW2015, SW2016 and SW2017, the least: SNAP and ReaxFF. A detailed look shows that only two potentials correctly reproduce cohesive energy. Mechanical stability is correctly reproduced by all potentials, i.e., $K_i > 0$. Potential ReaxFF again catastrophically badly reproduces mechanical properties of 1T-MoS$_2$, even the symmetry of the stiffness tensor is again not correct. Unfortunately, the three potentials: SW2013, SW2015, SW2016, do not correctly reproduce the symmetry of the 1T-MoS$_2$ phase, i.e., during pre-relaxation input 1T-MoS$_2$ converges to 1H-MoS$_2$ phase.

Table 3: Structural and mechanical properties of SL 1T-MoS$_2$ from molecular calculations: lattice parameters a,b (Å), average cohesive energy $E_c$ (eV/atom), average bond lengths $d$ (Å), average height $h$ (Å), 2D elastic constants $C_{ij}$ (N/m), 2D Kelvin moduli $K_i$ (N/m), mean absolute percentage error (MAPE) (%)

| Method | DFT | SW2013 | SW2015 | SW2016 | SW2017 | REBO | SNAP | ReaxFF |
|---|---|---|---|---|---|---|---|---|
| a | 3.194 | 3.062* | 3.117* | 3.174* | 3.307 | 3.194 | 3.072 | 3.162 |
| b | 3.194 | 3.062* | 3.117* | 3.174* | 3.307 | 3.194 | 3.072 | 3.162 |
| $-E_c$ | 5.52 | 3.00 | 0.62 | 1.84 | 4.96 | 7.05 | 2.31 | 4.84 |
| $d_{Mo-S}$ | 2.422 | 2.399 | 2.382 | 2.515 | 2.42 | 2.445 | 2.476 | 2.433 |
| $h_{S-S}$ | 3.142 | 4.223 | 4.257 | 4.032 | 2.973 | 3.211 | 3.454 | 3.203 |
| $C_{11}$ | 84.1 | 103.9 | 45.8 | 91.7 | 121.8 | 118.2 | 437.1 | 173.3 |
| $C_{22}$ | 84.1 | 103.9 | 45.8 | 91.7 | 121.8 | 118.2 | 437.1 | 32.1 |
| $C_{12}$ | 5.0 | 33.4 | 8.0 | 28.4 | 28.6 | 32.4 | 6.1 | 83.8 |
| $C_{44}$ | 39.6 | 35.2 | 18.9 | 31.7 | 46.6 | 42.9 | 215.5 | 9.4 |
| $K_I$ | 89.1 | 137.3 | 53.8 | 120.1 | 150.4 | 150.6 | 443.2 | 147.8 |
| $K_{II}$ | 79.1 | 70.5 | 37.8 | 63.3 | 93.2 | 85.8 | 431.0 | 57.6 |
| $K_{III}$ | 79.2 | 70.4 | 37.8 | 63.4 | 93.2 | 85.8 | 431.0 | 18.8 |
| MAPE $_{1T}$ | | 65.962 | 39.849 | 56.735 | 58.860 | 62.843 | 222.509 | 167.192 |

* Input 1T converges to 1H

Identified thirteen structural and mechanical properties of the SL 1T'-MoS$_2$ phase are summarised in Tab.4. In general, analysing the MAPE $_{1T'}$ for 1T'-MoS$_2$, the three most accurate potentials are: SW2016, REBO and SW2017, the least: SNAP and ReaxFF. Once again only two potentials correctly reproduce cohesive energy. Mechanical stability is reproduced in the right way by all potentials, i.e., $K_i > 0$. Unfortunately, the two potentials: SW2017 and SNAP, do not properly restore the symmetry of the 1T'-MoS$_2$ phase, i.e., during pre-relaxation input 1T'-MoS$_2$ basic cell converges to 1T-MoS$_2$.

Let us now analyse the cumulative performance of the analysed potentials for all SL MoS$_2$ phases. We see that MAPE in Tab.4 is the lowest, and almost the same, for three potentials: SW2017, SW2016 and REBO. However, only the REBO potential distinguishes three different phases, the other two potentials degenerate phases, i.e., instead of three they produce two.

Table 4: Structural and mechanical properties of SL 1T'-MoS$_2$ from molecular calculations: lattice parameters a,b (Å), average cohesive energy $E_c$ (eV/atom), average bond lengths $d$ (Å), average height $h$ (Å), 2D elastic constants $C_{ij}$ (N/m), 2D Kelvin moduli $K_i$ (N/m), mean absolute percentage error (MAPE) (%)

| Method | DFT | SW2013 | SW2015 | SW2016 | SW2017 | REBO | SNAP | ReaxFF |
|---|---|---|---|---|---|---|---|---|
| a | 5.751 | 4.944 | 5.757 | 5.263 | 5.728[†] | 5.563 | 5.321[†] | 5.609 |
| b | 3.177 | 3.062 | 3.148 | 3.172 | 3.307[†] | 3.245 | 3.072[†] | 3.209 |
| $-E_c$ | 5.56 | 3.02 | 0.55 | 1.87 | 4.96 | 6.93 | 2.31 | 4.83 |
| $d_{Mo-S}$[‡] | 2.415 | 2.399 | 2.406 | 2.504 | 2.42 | 2.468 | 2.476 | 2.490 |
| $h_{S-S}$ | 3.364 | 4.641 | 5.173 | 4.142 | 2.973 | 3.781 | 3.454 | 3.399 |
| $C_{11}$ | 68.1 | 1.1 | 0.0 | 60.4 | 121.8 | 56.8 | 437.1 | 120.1 |
| $C_{22}$ | 78.9 | 100.5 | 37.6 | 94.6 | 121.8 | 113.0 | 437.1 | 255.7 |
| $C_{12}$ | 18.2 | 1.1 | 0.0 | 20.3 | 28.6 | 23.1 | 6.1 | 68.1 |
| $C_{44}$ | 43.2 | 27.1 | 0.0 | 26.9 | 46.6 | 70.5 | 215.5 | 6.4 |
| $K_I$ | 90.9 | 100.5 | 37.6 | 88.4 | 150.4 | 121.3 | 443.2 | 194.3 |
| $K_{II}$ | 56.1 | 1.1 | 0.0 | 66.6 | 93.2 | 48.5 | 431.0 | 181.5 |
| $K_{III}$ | 86.4 | 54.2 | 0.0 | 53.8 | 93.2 | 141.0 | 431.0 | 12.8 |
| MAPE $_{1T'}$ | | 42.070 | 63.020 | 20.110 | 30.399 | 25.395 | 249.177 | 91.913 |
| L$_{MAPE}$ | | 127.830 | 151.074 | 102.187 | 100.522 | 104.840 | 483.573 | 325.504 |

[†] Input 1T' converges to 1T
[‡] average first-neighbour bond lengths calculated with *cutoff* radius=3.5 and number of histogram bins=50

*3.2. Phonon spectra*

Phonon spectra along the appropriate high symmetry q-points [37], calculated by applying the PBEsol XC functional, for SL 1H-MoS$_2$ phase are depicted in Fig.4a), for SL 1T-MoS$_2$ phase are depicted in Fig.4b) and for SL 1T'-MoS$_2$ phase are shown in Fig.4c), respectively. Experimental data for single-layer molybdenum disulphide are very scarce and concern only $\Gamma$ point in 1H-MoS$_2$ phase, see [58]. When we compare the results obtained here with those calculated by other authors, we see agreement typical for different DFT calculations, see [59, 60, 61, 62, 31].

Analysis of the computed curves in Figs.4 a),b),c) allows us to conclude that phases 1H-MoS$_2$ and 1T'-MoS$_2$ are not only mechanically but also dynamically stable, i.e., phonon modes everywhere have positive frequencies. Phase 1T-MoS$_2$ is mechanically stable, but not dynamically stable, i.e., phonon modes also have negative frequencies. Similar observations can be found in [31, 60].

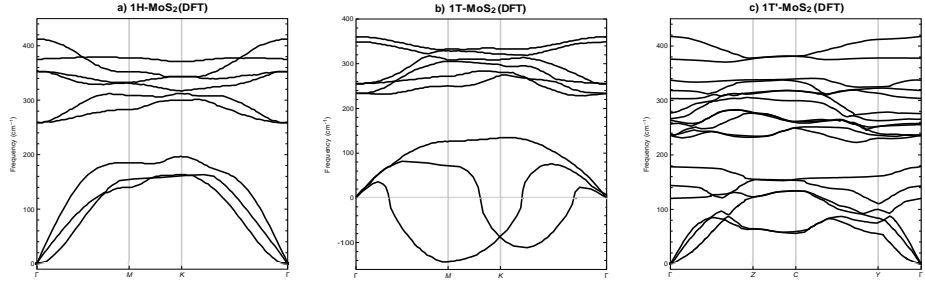

Figure 4: Phonon dispersion of SL MoS$_2$ from DFT **a)** 1H, **b)** 1T and **c)** 1T′. High symmetry points: Γ [0,0,0], **M**[1/2,0,0], **K**[1/3,1/3,0], **Z**[0,1/2,0], **C**[1/2,1/2,0], **Y**[1/2,0,0].

Let us now compare the phonon spectra for SL MoS$_2$ phases calculated with DFT, Figs.4, and these calculated with LAMMPS and the three best potentials, i.e., SW2017, Figs.5, SW2016, Figs.6 and REBO, Figs.7. Because only the REBO potential distinguishes three different phases of SL MoS$_2$ the molecular calculations of phonons utilises basic cells derived from DFT calculations. At first glance, we can see that it only makes sense to compare molecular phonons with DFT phonons merely qualitatively, not quantitatively. All three potentials are qualitatively well reproducing the phonon spectra for SL 1H-MoS$_2$ phase, see Fig.5a), Fig.6a) and Fig.7a). For SL 1T-MoS$_2$ phase, only SW2016 potential predicts dynamical instability, see Fig.6b). For SL 1T'-MoS$_2$ phase, SW2017 and REBO potentials behave reasonably, see Fig.5c) and Fig.7c). The conclusion is that none of the three potentials correctly reproduces the dynamical stability of all SL MoS$_2$ phases.

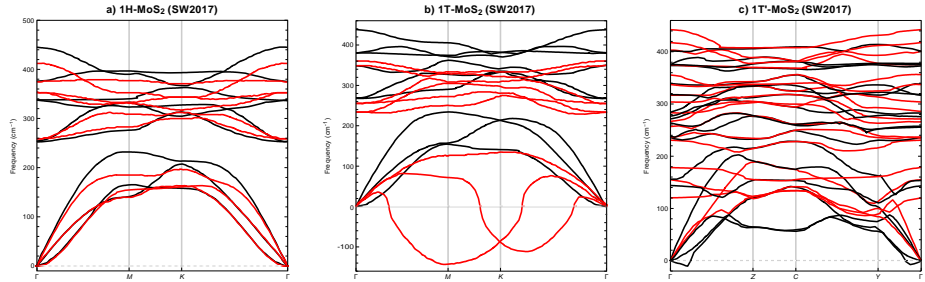

Figure 5: Phonon dispersion of SL MoS$_2$ from SW2017 potential **a)** 1H, **b)** 1T and **c)** 1T′. Black lines represent SW2017 results, red lines represent DFT results.

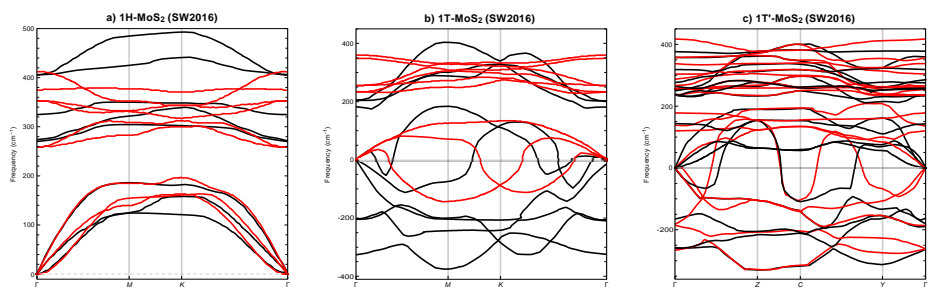

Figure 6: Phonon dispersion of SL MoS$_2$ from SW2016 potential **a**) 1H, **b**) 1T and **c**) 1T'. Black lines represent SW2016 results, red lines represent DFT results.

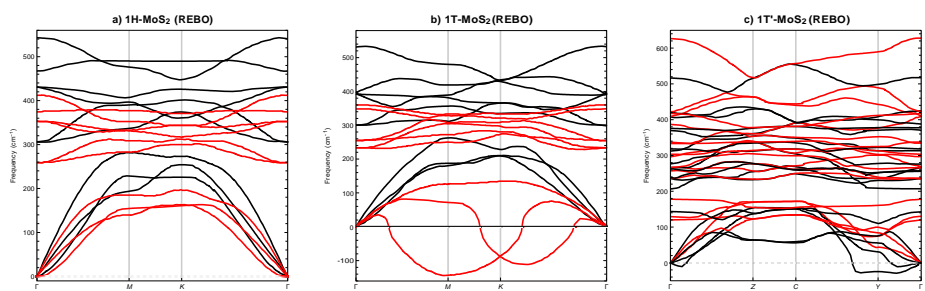

Figure 7: Phonon dispersion of SL MoS$_2$ from REBO potential **a**) 1H, **b**) 1T and **c**) 1T'. Black lines represent REBO results, red lines represent DFT results.

## 4. Conclusions

A systematic quantitative comparison of Stillinger-Weber, REBO, SNAP and ReaxFF potentials for the reproduction of the properties of 2D molybdenum disulphide polymorphs was presented. To compare the potentials, the structural and mechanical properties and phononon dispersion of single-layer phases 2H, 1T and 1T' MoS$_2$ (SL MoS$_2$) obtained from the functional density theory (DFT) and molecular static (MS) calculations were used.

We can conclude that:

- The transferability of analysed molecular potentials leaves much to be desired.

- Three potentials: SW2016, SW2017 and REBO demonstrate the best quantitative performance.

- None of the above three potentials correctly reproduces the dynamical stability of all SL MoS$_2$ phases.

- Only the REBO potential distinguishes three different 2D molybdenum disulphide allotropes.

- Two potentials: ReaxFF and SNAP demonstrate significantly lower quantitative efficiency.

- It seems that the low transferability of the analysed potentials is a result of the improper fitting of their parameters.

- To increase the transferability of potentials, the number of configurations to be taken into account in the parameter optimisation process should be significantly increased.

I hope that the observations made here will help other researchers to choose the right potentials for their purposes and will be a suggestion for parametrising new potentials for SL $MoS_2$.

**ACKNOWLEDGMENTS**

This work was supported by the National Science Centre (NCN – Poland) Research Project: No. 2016/21/B/ST8/02450. Additional assistance was granted through the computing cluster GRAFEN at Biocentrum Ochota, the Interdisciplinary Centre for Mathematical and Computational Modelling of Warsaw University (ICM UW) and Poznań Supercomputing and Networking Center (PSNC).

**Appendix A. Supplementary Materials**

```
# CIF file 1H-MoS2
# This file was generated by FINDSYM
# Harold T. Stokes, Branton J. Campbell, Dorian M. Hatch
# Brigham Young University, Provo, Utah, USA

data_findsym-output
_audit_creation_method FINDSYM

_symmetry_space_group_name_H-M "P -6 m 2"
_symmetry_Int_Tables_number 187

_cell_length_a   3.16544
_cell_length_b   3.16544
_cell_length_c   16.00000
_cell_angle_alpha  90.00000
_cell_angle_beta  90.00000
_cell_angle_gamma 120.00000

loop_
_space_group_symop_id
_space_group_symop_operation_xyz
```

```
1 x,y,z
2 -y,x-y,z
3 -x+y,-x,z
4 x,x-y,-z
5 -x+y,y,-z
6 -y,-x,-z
7 -x+y,-x,-z
8 x,y,-z
9 -y,x-y,-z
10 -x+y,y,z
11 -y,-x,z
12 x,x-y,z

loop_
_atom_site_label
_atom_site_type_symbol
_atom_site_symmetry_multiplicity
_atom_site_Wyckoff_label
_atom_site_fract_x
_atom_site_fract_y
_atom_site_fract_z
_atom_site_occupancy
Mo1 Mo   1 d 0.33333 0.66667 0.50000 1.00000
S1  S    2 i 0.66667 0.33333 0.40249 1.00000

# CIF file 1T-MoS2
# This file was generated by FINDSYM
# Harold T. Stokes, Branton J. Campbell, Dorian M. Hatch
# Brigham Young University, Provo, Utah, USA

data_findsym-output
_audit_creation_method FINDSYM

_symmetry_space_group_name_H-M "P -3 2/m 1"
_symmetry_Int_Tables_number 164

_cell_length_a    3.19358
_cell_length_b    3.19358
_cell_length_c    16.00000
_cell_angle_alpha 90.00000
_cell_angle_beta  90.00000
_cell_angle_gamma 120.00000

loop_
_space_group_symop_id
_space_group_symop_operation_xyz
```

```
1 x,y,z
2 -y,x-y,z
3 -x+y,-x,z
4 x-y,-y,-z
5 y,x,-z
6 -x,-x+y,-z
7 -x,-y,-z
8 y,-x+y,-z
9 x-y,x,-z
10 -x+y,y,z
11 -y,-x,z
12 x,x-y,z

loop_
_atom_site_label
_atom_site_type_symbol
_atom_site_symmetry_multiplicity
_atom_site_Wyckoff_label
_atom_site_fract_x
_atom_site_fract_y
_atom_site_fract_z
_atom_site_occupancy
Mo1 Mo   1 b 0.00000 0.00000 0.50000 1.00000
S1   S    2 d 0.33333 0.66667 0.40182 1.00000

# CIF file 1T'-MoS2
# This file was generated by FINDSYM
# Harold T. Stokes, Branton J. Campbell, Dorian M. Hatch
# Brigham Young University, Provo, Utah, USA

data_findsym-output
_audit_creation_method FINDSYM

_symmetry_space_group_name_H-M "P 1 21/m 1"
_symmetry_Int_Tables_number 11

_cell_length_a   5.75123
_cell_length_b   3.17711
_cell_length_c   16.00000
_cell_angle_alpha  90.00000
_cell_angle_beta  90.00000
_cell_angle_gamma  90.00000

loop_
_space_group_symop_id
_space_group_symop_operation_xyz
```

```
1 x,y,z
2 -x,y+1/2,-z
3 -x,-y,-z
4 x,-y+1/2,z

loop_
_atom_site_label
_atom_site_type_symbol
_atom_site_symmetry_multiplicity
_atom_site_Wyckoff_label
_atom_site_fract_x
_atom_site_fract_y
_atom_site_fract_z
_atom_site_occupancy
Mo1  Mo    2 e 0.70568 0.25000 0.49753 1.00000
S1   S     2 e 0.41977 0.25000 0.60514 1.00000
S2   S     2 e 0.07725 0.25000 0.41677 1.00000
```